\documentclass[pra,a4paper,twocolumn,showpacs,amsmath,amssymb,floatfix,amstex]{revtex4}

\usepackage[dvips]{graphicx}
\usepackage{bm,pstricks,longtable}

\input{epsf}
\newcommand{\ket}[1]{| #1\rangle}                       %

\newcommand{\equa}[1]{Eq.~(\ref{#1})}  

\def\nq{n}

\begin{document}

\title{Quantum computation of multifractal exponents through the quantum wavelet transform}
\author{Ignacio Garc\'ia-Mata$^{1,2}$, Olivier Giraud$^{1,2}$ 
and Bertrand Georgeot$^{1,2}$}
\affiliation{$^1$  Universit\'e de Toulouse; UPS; Laboratoire de Physique 
 Th\'eorique (IRSAMC);  F-31062 Toulouse, France \\
$^2$ CNRS; LPT (IRSAMC); F-31062 Toulouse, France}

\begin{abstract}
We study the use of the quantum wavelet transform to extract efficiently
information about the multifractal exponents for multifractal quantum
states.  We show that, combined with quantum 
simulation algorithms, it enables to build
quantum algorithms for multifractal exponents with a polynomial gain compared
to classical simulations.  Numerical results indicate
that a rough estimate of fractality could be obtained exponentially fast.
Our findings are relevant e.g. 
for quantum simulations of multifractal quantum maps
and of the Anderson model at the metal-insulator transition.
\end{abstract}
\date{December 31, 2008}
\maketitle
\section{Introduction}

It has been realized over the past twenty years
that the specific properties
of quantum mechanics enable to conceive new ways of treating 
and manipulating information (for a review see e.g.
\cite{nielsen}).  In particular, the idea
of a quantum computer has been put forth.  Such a device would
perform computation on quantum registers usually thought of as made
of qubits.  It has been shown that quantum algorithms can be devised
which are asymptotically faster than classical algorithms, the most
famous examples being the Shor algorithm which factorizes numbers 
exponentially faster than any known classical algorithm \cite{shor}, 
and the Grover
algorithm which searches a database quadratically faster than any
possible classical procedure \cite{grover}.

However, not so many efficient quantum algorithms have been found, and
it is still unclear which problems can be treated faster on a quantum
computer and how efficiently.  One of the first possibilities to be put
forward was the simulation of quantum systems on quantum computers, which was 
first envisioned by Feynman \cite{feynman} and then made more precise
in several subsequent works 
\cite{lloyd,abrams,schack,georshe,saw1,georlevshe,georlev}.  Some 
of these algorithms have been experimentally implemented in few-qubit systems
such as in \cite{cory}. Recent more mathematical works have established
more rigorously that a quantum computer can indeed simulate efficiently
(i.e. exponentially fast)
a wide class of quantum systems \cite{israeli,sanders}. However,
these works were mostly 
concerned with the possibility of efficient simulation of
a quantum system. To be complete and comparable with a classical
algorithm, a quantum algorithm should 
not only perform a computation in an efficient way, but also be able
to extract information from the result of this computation in an efficient 
way. Thus a quantum algorithm should also include specification of 
how information is extracted from the final wavefunction at the end of the 
simulation.  Several proposals have been made in order to extract
efficiently information from a quantum simulation, looking at 
the fidelity \cite{fidelity}, the spectral statistics \cite{formfactor},
the localization length \cite{loclength}, 
the Wigner function \cite{georlevshe}, or the diffusion constants 
\cite{como,georlev}. 
It has been found that the final gain compared to classical simulation 
depends on the
choice of the observable and on the measurement procedure, and can
dramatically change the efficiency of the quantum simulation.  It is therefore
important to explore in more detail which quantities can be extracted from
a quantum simulation and with which efficiency.

The quantum systems studied up to now in this setting 
were in general localized or extended in the computational basis.  
However, there exists another class
of quantum systems intermediate between these two types, whose 
wavefunctions are multifractal.  Such properties appear
in physical systems, for example in wavefunctions of
 electrons in a disordered potential
at the Anderson transition between metal and insulator 
\cite{mirlin,cuevas,garciagarcia,fyodorov}, 
or at the quantum Hall transition \cite{huckestein}.  It has been shown 
that simple
systems displaying such properties can be simulated exponentially
fast on a quantum computer \cite{pomer,interm}.  It is therefore interesting 
to assess if the multifractal properties can be extracted from
the resulting wavefunction at the end of the quantum simulation, and
with which efficiency.

In this paper, we explore different strategies to measure the multifractality,
and more precisely multifractal exponents,
from a quantum wavefunction produced by a quantum simulation, and assess 
their efficiency.  In particular, we assess the interest of the 
wavelet transform to perform such tasks.  This transform is a generalization
of the Fourier transform using basis functions localized in both position
and momentum instead of the sinusoidal waves of the Fourier transform.
The wavelet transform has been used with great success in data treatment and 
data compression, and has been included in compression standards such as 
MPEG4.  It has been shown that a quantum wavelet transform (QWT) can be
implemented efficiently on a quantum computer \cite{WT1,WT2,WT3}, but
it has been seldom used in quantum algorithms, with few exceptions 
\cite{wavealg2,terraneo,husimi,wavealg1,wavealg}.  
As concerns multifractal analysis, recent theoretical progress in classical
data treatment have shown that the wavelet 
transform can be used as a versatile tool to explore multifractality of 
various phenomena, such as DNA sequences, turbulence  or
cloud structure \cite{arneodo}.

Our results show that, although the methods of classical multifractal
analysis cannot be directly implemented efficiently on a quantum computer, 
suitable modifications of the method combined with amplitude
amplification enable to get multifractal
coefficients from a wavefunction with polynomial efficiency.  
This translates into complete algorithms of quantum simulations including
measurements which are polynomially faster than corresponding
classical algorithms.  We also give numerical indications that
the wavelet transform may enable to assess a rough measure of the 
degree of fractality of a wavefunction with exponential efficiency.  

\section{Multifractal quantum states}

Multifractal quantum states are characteristic of certain systems
intermediate between chaos and integrability.  They have been found
at the Anderson transition between metal and insulator, i.e. between
extended states and localized states, and also at
quantum Hall transitions \cite{mirlin,cuevas,garciagarcia,fyodorov,huckestein}.
  More generally, they correspond to a 
whole class of systems whose spectral statistics, for example, are
of the semi-poisson type \cite{gerland} 
intermediate between Random Matrix statistics
(associated with ergodic states) and Poisson statistics (associated 
with integrable or localized states).

Multifractality properties of wavefunctions are described by a
whole set of generalized fractal dimensions $D_q$. For a vector
$|\psi\rangle=\sum_{i=1}^{N}\psi_i|i\rangle$ in an $N$-dimensional
Hilbert space, the spectrum $\tau_q$ is defined through
the scaling of the moments
\begin{equation}
\label{dimfrac}
\sum_{i=1}^N |\psi_i|^{2q} \propto N^{-\tau_q}.
\end{equation}
The multifractal exponents, or generalized multifractal dimensions,
 are related to the spectrum by the relation
\begin{equation}
D_q=\frac{\tau_q}{q-1}.
\end{equation}
In order to investigate properties of multifractal quantum states,
simple models have been devised which exhibit such properties 
\cite{giraud,bogomolny,garciagarcia}.
A particularly simple example is the quantization of a map
 on the torus which has been shown to 
exhibit a whole range of multifractal properties, depending on one parameter. 
The classical map is given by
\begin{equation}
\begin{array}{cccc}
\bar{p} &=& p + \gamma &\mathrm{(mod} \;\mathrm{1)}\\
\bar{q} &=& q+ 2\bar{p} &\mathrm{(mod} \;\mathrm{1)}
\end{array}
.
\end{equation}
The quantization of this map yields
a unitary evolution operator acting on a Hilbert space of dimension
$N=1/(2\pi\hbar)$ which can be expressed in momentum space by the $N
\times N$ matrix~\cite{giraud,bogomolny}
\begin{equation}\label{ISRM}
    U_{pp'}=\frac{e^{i\phi_p} }{N}\frac{1-e^{2i\pi N
\gamma}}{1-e^{2i\pi (p-p'+N\gamma)/N}},
\end{equation}
with $\phi_p=-2\pi p^2/N$. 
From this quantized map one can
construct an ensemble of random matrices, taking $\phi_p$ as independent
random variables uniformly distributed in $[0, 2\pi[$ \cite{bogomolny}. 

This map has different properties depending on the parameter
$\gamma$.  Indeed, for irrational $\gamma$, the map possesses
the characteristics of systems displaying quantum chaos, with
eigenstates ergodic over the phase space and spectral statistics 
of the eigenphases of the evolution operator 
following random matrix predictions.  In contrast, 
for rational $\gamma=n_1/n_2$, the spectral statistics 
are of the semi-Poisson type intermediate between those of
integrable and chaotic systems \cite{bogomolny}.
The eigenstates in momentum representation display
multifractal properties studied in \cite{martin}.
The fractality depends on $n_2$ and is stronger for smaller $n_2$. 
Thus the set of quantum maps $U$ with rational $\gamma$
gives a random
matrix ensemble with intermediate statistics (ISRM) 
whose multifractal properties are controlled by the parameter
$n_2$.  Such a map can be implemented efficiently on a quantum computer
\cite{interm}, as an exponentially large vector can be evolved
through (\ref{ISRM}) with a polynomial number of gates. 
This map will be used as a benchmark for numerical simulations
in this paper, since it represents a simple but non-trivial example
of system with multifractal properties depending on one parameter
and which can be simulated efficiently on a quantum computer. 

On the other hand, this system is complicated enough to be
hard to tackle analytically.
In order to test the accuracy of our approach for the 
estimation of multifractal dimensions on a simpler system
where all exponents can be analytically computed, we consider the
situation where the coefficients of the quantum state are 
given by a multifractal cascade. Multifractal cascades
are examples of multifractal measures for which generalized
dimensions can be calculated analytically. The simplest example
was proposed in \cite{cascade}; it is a special case of
a Bernoulli measure on a two-scale Cantor set \cite{procaccia}.
It can be constructed by the following process:
one breaks an initial interval into two equal parts,
attributes a weight $p_1$ to one
half and $p_2=1-p_1$ to the other, and
repeats the process (keeping the same weights $(p_1,p_2)$ constant)
on the newly constructed intervals. After $k$ steps, there are $2^k$
intervals and the weight of each interval is of the form $p_1^r p_2^{k-r}$
for some $r$, $0\leq r\leq k$. The generalized dimensions can be shown 
to be given by \cite{procaccia}
\begin{equation}
\label{Dq_cascade}
D_q=\frac{1}{1-q}\log_2\left(p_1^q+p_2^q\right).
\end{equation}
In this way one can construct quantum 
``cascade'' states of size $2^k$ whose amplitudes squared are 
given by the weights of the cascade.

\section{Wavelet transforms and multifractal properties}
An important procedure has been developed in recent years to 
extract multifractal exponents from a distribution. 
It uses the wavelet transform 
\cite{waveletbook}, a generalization of the Fourier transform which 
expands a function on the wavelet basis instead of the Fourier basis.
Contrary to the sinusoidal waves which compose the Fourier basis, 
which have specified frequencies but are extended in position space,
the wavelets are localized both in momentum and position.  They can thus 
probe many properties which are difficult to reach with the standard Fourier
transform, such as singularities of the distribution.
This has made wavelet transforms a popular tool in recent developments
of e.g. image or sound treatment and compression in classical
information, such as the formats JPEG and MPEG.

Wavelet transforms are based on a single function $g$, called the
analyzing wavelet or mother wavelet. 
The wavelet transform of a function $f$ is a function $T_f$ of two
variables defined as 
\begin{equation}
\label{wave}
T_f(a,b)=\frac{1}{a}\int dx\, f(x) g\left(\frac{x-b}{a}\right).
\end{equation}
Variable $a$ corresponds to the scale at which the function $f$ is
analyzed, while $b$ is a space variable. Thus $T_f(a,b)$ is a measure
of how close the function $f$ is to the mother wavelet at point $b$
and at scale $a$.

If the function $f$ is sampled as a $N$-dimensional vector where $N=2^n$, the
wavelet transform can be discretized and implemented as a unitary 
transformation, resulting in a Fast Wavelet Transform (FWT)
analogous to the Fast Fourier Transform. The scale parameter $a$  
takes values $1,1/2,1/4,...1/2^{n-1}$, while the space parameter $b$ 
varies over ${\cal L}(a)=\{1,2,\ldots,1/a\}$ at scale $a$. 
A discrete version of the
mother wavelet is constructed recursively at each scale.
Commonly used mother wavelets for the FWT include the Haar 
wavelet \cite{Haar} and the Daubechies wavelet \cite{daub}.  Throughout
the paper, we will use the Daubechies 4 discrete wavelet transform 
in the numerical simulations.
 It has been shown that a quantum wavelet transform (QWT) implementing
such discrete wavelet transforms can be performed efficiently on a quantum 
computer \cite{WT1,WT2,WT3}, namely
an exponentially large vector $\sum_{i=0}^{N-1}\psi_i\ket{i}$ of size 
$N=2^{\nq}$ can be
transformed in a number of operations polynomial in $\nq$ into a vector
$\sum_{a,b}T_{\psi}(a,b)\ket{a,b}$, where the wavelet transform 
at scale $a$ is stored on the computational basis vectors $\ket{a,.}$.
\begin{figure}[htbp]
\begin{center}
\includegraphics[width=1.0\linewidth]{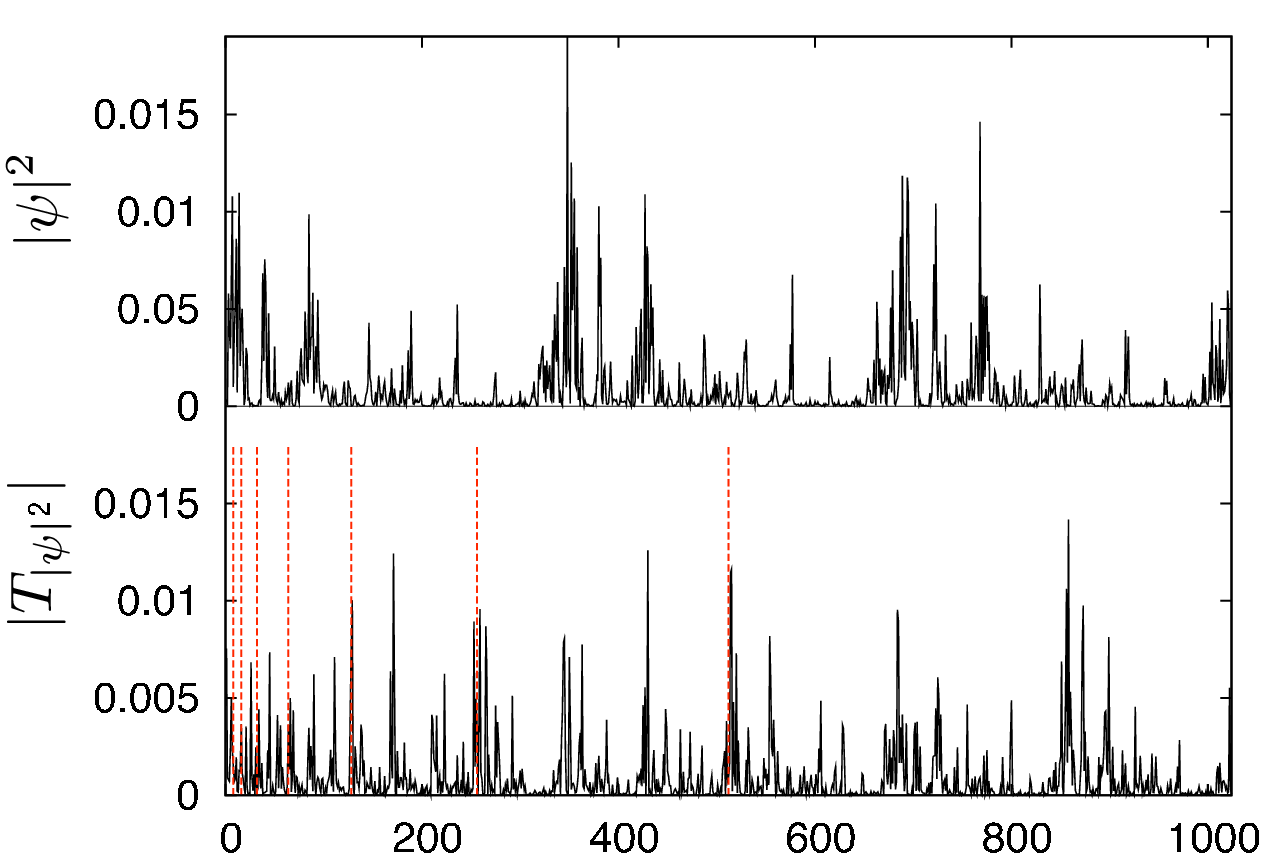}
\caption{(Color online) Example of one eigenfunction of (\ref{ISRM}) (top) 
and its wavelet transform (bottom).  
The red/grey dashed vertical lines on the bottom panel
separate the different scales in the wavelet transform.}
\label{fig1}
\end{center}
\end{figure}

The wavelet transform has been put forward recently as a tool for
extracting the value of the exponents from a multifractal
distribution.
These exponents  are usually quite hard to extract numerically,
and are very unstable, since the values obtained depend on the 
chosen numerical method  
up to very large system sizes.  An example of a multifractal
distribution (for an eigenvector of (\ref{ISRM})) is shown on Fig.\ref{fig1},
together with its wavelet transform.  One sees that the wavelet
transform does not look especially simpler than the original distribution.
However, recent works have shown \cite{arneodo,arnschol,arnbook}
that the wavelet transform allows to extract the exponents of the
distribution $f$, using the maxima of the wavelet transform at each
scale.  
Such methods based on the wavelet transform have enabled to extract 
multifractal exponents in complicated physical systems of great
fundamental and technological importance such as e.g. 
DNA sequences \cite{arne2}, fully developed turbulence \cite{arne3} or
high-resolution satellite images of cloud structure \cite{arne4}.

However, these methods use the continuous wavelet transform, which
is delicate to implement for many systems of interest.
A variation of the wavelet method developed in
\cite{allemands,coreens}
uses instead of the maxima of the continuous wavelet transform the
sum of the values of the discrete wavelet transform,
properly normalized at each scale. One defines the partition function
 \begin{equation}
\label{Zarneoff}
Z(a,q)=\sum_{b\in {\cal L}(a)}\left[\frac{|T_f(a,b)|}
{\sum _{b\in {\cal L}(a)}|T_f(a,b)|}\right]^q
\end{equation}
where $a$ is the scale  and ${\cal L}(a)$ is the interval corresponding to
each scale $a$. The asymptotic behavior of the partition function at small scales
is governed by the generalized dimensions as
\begin{equation}
\label{eq:tauq}
Z(a,q) \sim_{a \rightarrow 0^+} a^{\tau_q}
\end{equation}
with $\tau_q \equiv D_q(q-1)$.

\section{quantum algorithms for multifractal exponents}

There does not seem to be a simple way to implement efficiently
the maxima method on a quantum computer.
However, as we show below, 
the scaling \eqref{eq:tauq} of the partition function (\ref{Zarneoff})
can be evaluated
on a quantum computer, and gives rise to quantum algorithms for 
estimating multifractal exponents $\tau_q$.
Our goal in this section is to construct quantum algorithms
which use Eqs.~\eqref{Zarneoff}-\eqref{eq:tauq}
in order to efficiently extract information about multifractal
properties, in the situation where
quantum states are obtained by quantum simulation.

\subsection{Wavelet transform of $|\psi|^2$}
For a  given state $\psi$, the weights $|\psi_i|^2/\sum_i|\psi_i|^2$
define a normalized measure on the unit interval.
Following Eq.~\eqref{Zarneoff}, the partition function that describes 
the multifractal properties of $\psi$ reads
\begin{equation}
\label{Zarne}
Z_{|\psi|^2}(a,q)= \sum_{b\in {\cal L}(a)}\left[\frac{|T_{|\psi|^2}(a,b)|}{\sum _{b\in {\cal L}(a)}|T_{|\psi|^2}(a,b)|}\right]^q,
\end{equation}
where as before 
$a$ is the scale and ${\cal L}(a)$ is the interval corresponding 
to each scale $a$.
The denominator in \equa{Zarne} is a normalization that ensures that
at each scale 
the sum yields a proper probability measure.
The multifractal exponents $\tau_q$ can be extracted
in the limit of small scales through the scaling \eqref{eq:tauq}.
\begin{figure}[htbp]
\begin{center}
\includegraphics[width=1.0\linewidth]{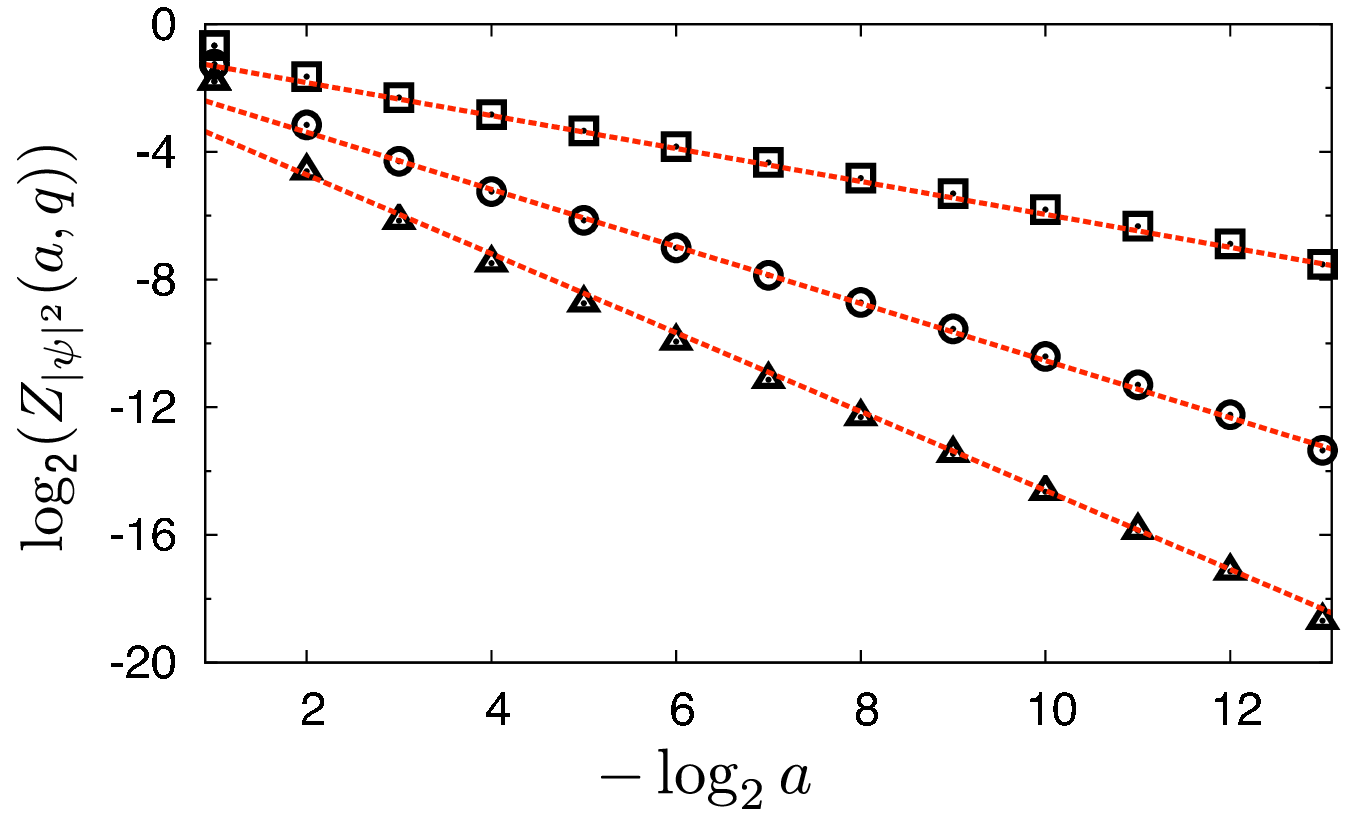}
\caption{(Color online) Average of $\log_2 Z_{|\psi|^2}(a,q)$ 
as a function of the scale 
$a$ for $q=2$ ($\square$), $q=3$ ($\circ$), $q=4$ ($\triangle$). The  
slope of the fitting straight lines (dashed) gives  
$-\tau_q$. Averaging is done over $\sim 25000$
eigenvectors of \eqref{ISRM} of size $N=2^{14}$
for $n_2=3$ ($\gamma=1/3$).}
\label{fig2}
\end{center}
\end{figure}
As is shown in Fig.\ref{fig2} for eigenvectors of \eqref{ISRM},
the partition function (\ref{Zarne}) indeed
scales as predicted for multifractal wavefunctions, 
and gives the
multifractal exponent $\tau_q$, whence the fractal dimensions.
For instance the linear fit for $q=2$ has a slope giving $\tau_2=D_2=0.52$,
which is consistent with the fractal dimension obtained in \cite{martin}.  

We now show that it is possible to implement such sums on a quantum 
computer in the case $q=2$, allowing to obtain the multifractal exponent $\tau_2$
for wavefunctions produced by a quantum simulation algorithm.
The first step is to adapt such a quantum simulation algorithm
to our present purpose. Usually the vector $\ket{\psi}$ we are interested 
in comes out of the quantum simulation with its components $\psi_i$
as amplitudes of the basis states. However in \eqref{Zarne} the wavelet 
transform is applied to a vector whose components are $|\psi_i|^2$.
Thus one first has to modify the quantum simulation
in order to produce $\sum_i |\psi_i|^2 |i\rangle$
rather than $\sum_i \psi_i |i\rangle$. To do so, one should build 
the product of two iterates on separate registers, in order to obtain a state
$\sum_{i=0}^{N-1}\sum_{j=0}^{N-1} \psi_i\psi^{*}_j|i\rangle
|j\rangle$, and then use amplitude amplification to keep only the
diagonal terms $i=j$. Let us detail the steps.
We start with two copies of an initial state 
$|\psi_0\rangle=\sum_{i=0}^{N-1} \psi^0_i |p_i\rangle $ in momentum
representation: $|\psi_0\rangle \otimes |\psi^*_0 \rangle
=\sum_{i=0}^{N-1}\sum_{j=0}^{N-1} \psi^0_i\psi^{0*}_j|p_i\rangle
|p_j\rangle$. Here $\psi_0^*$ is the complex conjugate of $\psi_0$
and $\ket{p_i}$ are the basis vectors in the momentum representation.
This step requires $2\nq$ qubits to hold the values of the wavefunction 
on a $N$-dimensional Hilbert space, where $N=2^{\nq}$.
Let us consider the iterations of such a vector through a quantum map.  
We can apply the algorithm implementing the evolution operator 
of the intermediate map $\hat{U}$ as explained in \cite{interm}
to each subsystem independently.
The operator on one subsystem can be described as the product of 
diagonal operators followed by Fourier transforms. On the quantum registers
this corresponds to multiplication by phases followed by 
quantum Fourier transform (QFT).
The multiplication by phases of each coefficient keeps the separability of
the state into its two subsystems. The QFT mixes only states with the same 
value of the other register attached, and therefore also keeps the factorized form.  
Let us see how this works for one iteration:
multiplication of the first register 
by the diagonal operator $e^{-i\phi_{p_i}}\delta_{ij}$
performs the transformation
\begin{equation}
\sum_{i=0}^{N-1}\sum_{j=0}^{N-1} 
\psi^0_i\psi^{0*}_j|p_i\rangle  |p_j\rangle
\rightarrow \sum_{i=0}^{N-1}\sum_{j=0}^{N-1} 
e^{-i\phi_{p_i}}\psi^0_i\psi^{0*}_j|p_i\rangle  |p_j\rangle.
\end{equation}
After QFT with respect to $p_i$ followed by multiplication by
$e^{2i\pi\gamma \hat{q}}$ the state can be put under the form
\begin{equation}
\sum_{j=0}^{N-1}(\sum_{i=0}^{N-1}
b_i |q_i\rangle) \psi^{0*}_j |p_j\rangle.
\end{equation}
Under QFT with respect to $q_i$ we get
\begin{eqnarray}
\sum_{j=0}^{N-1}(\sum_{i=0}^{N-1}
\psi^1_i |p_i\rangle)\psi^{0*}_j |p_j\rangle
&=&(\sum_{i=0}^{N-1}
\psi^1_i |p_i\rangle)\otimes (\sum_{j=0}^{N-1} \psi^{0*}_j |p_j\rangle)\nonumber\\
&=&\hat{U}|\psi_0\rangle \otimes |\psi^*_0 \rangle.
\end{eqnarray}
Thus iterations can be carried independently on each register.
By applying the same steps $2t$ times, we obtain the state
$\hat{U}^t |\psi_0\rangle \otimes \hat{U}^{*t} |\psi^*_0 \rangle
=\sum_{i=0}^{N-1}\sum_{j=0}^{N-1}
 \psi^t_i |p_i\rangle \otimes \psi^{t*}_j |p_j\rangle$.
This can be done in $O(t\nq^2)$ gates if we use the algorithm of 
\cite{interm} to implement $\hat{U}$.
Once this is done, the state after $t$ iterations is a product of two copies
of the state $\ket{\psi}$ whose fractal dimensions we are looking for.
In the momentum representation it takes the form
$\sum_{i,j}\psi_i \psi^{*}_j |p_i\rangle \otimes|p_j\rangle$.
We wish to select in this double sum the terms with $i=j$. This can be
done through amplitude amplification \cite{amplification},
which is a generalization of Grover's algorithm \cite{grover}.  The
latter starts from an equal superposition of $N$ states, and in $\sqrt{N}$
operations brings the amplitude of a specific state close to one.
Amplitude amplification increases the amplitude of a {\em whole subspace}.
If $P$ is a projector onto this subspace, and $\hat{V}$ is the operator
taking $|0\rangle$ to a state having a nonzero projection on the desired
subspace, repeated iterations of 
$\hat{V}(I-2|0\rangle\langle0|)\hat{V}^{-1}(I-2P)$ on $\hat{V} |0\rangle$
will increase the projection.  Indeed, if one writes $\hat{V} |0\rangle
=P\hat{V} |0\rangle +(I-P)\hat{V} |0\rangle$, the result of one iteration is to
rotate the state toward $P\hat{V} |0\rangle$ staying in the subspace 
spanned by $P\hat{V} |0\rangle$ and $(I-P)\hat{V} |0\rangle$. If 
$x=|P\hat{V} |0\rangle|^2$, one can check that after one iteration the 
state is $(4 x^2-3)P\hat{V} |0\rangle +(4 x^2-1)(I-P)\hat{V} |0\rangle$,
with a component along $(I-P)\hat{V} |0\rangle$ decreased by $4x^2$.

If $\hat{V}$ is chosen to be $\hat{U}^t \otimes \hat{U}^{*t} $ and $P$
to be a projector on the space corresponding to $i=j$,
the process of amplitude amplification from the initial state
$\ket{\psi_0}\otimes\ket{\psi^*_0}$ will bring the probability
of $i=j$ to one. The number of iterations
depends on the probability inside this space compared to total probability.
As we want to select terms with $p_i=p_j$ the subspace
we are searching for has a relative weight $\sum_{i=0}^{N-1} |\psi_i|^4$.
For a fractal state this quantity scales as $N^{-\alpha}$, 
with $\alpha \leq 1$.
Thus the total number of Grover iterations will be of the order $N^{\alpha/2}$.
So in the worst case of $\alpha\approx 1$, the total cost of building 
the state  $\sum_{i=0}^{N-1} |\psi_i|^2\ket{p_i}\ket{p_i}$ is 
$\sim t\sqrt{N}$. The more fractal the system is, the smaller $\alpha$ is, 
and the more efficient the algorithm is.

Once the state $\sum_{i=0}^{N-1} |\psi_i|^2\ket{p_i}\ket{p_i}$ 
is built, the second register is set back to $\ket{0}$ and
the QWT can be applied at a logarithmic cost on the first register. 
This yields 
a state $\sum_{a,b} T_{|\psi|^2}(a,b)|a,b\rangle$.
Then measurement of $a$ will give an histogram of 
$\sum_{b} |T_{|\Psi|^2}(a,b) |^2$
for the different values of $a$.
The denominator of (\ref{Zarne})
can be obtained through classical approximations
with lower resolution (e.g. half the number of qubits), 
and the exponent can be obtained from the slope
of $\log_2 Z_{|\psi|^2}(a,q)$.
The total cost for applying $t$ iterations of the map and extracting
the fractal dimension is thus $O(tN^{\alpha/2})$ operations, as opposed
to $tN$ for the classical algorithm.

Here we considered vectors obtained by iterations of a map. It
is however important to study also eigenvectors.
The procedure above cannot be
generalized directly, since eigenvectors
are to be selected by phase estimation first before the partition function
can be calculated.  

In order to obtain eigenvectors of a map $U$, one first builds the state 
$2^{-\nq}\sum_t |t\rangle|U^t\psi_0\rangle 
\sum_t' |t'\rangle|U^{*t'}\psi^*_0\rangle$ with 
$0 \leq t \leq N-1$.  
The initial state $|\psi_0\rangle$
should be simple enough to be built efficiently (for example
$|00...0\rangle$). To obtain the
double sum one starts from $2^{-\nq}\sum_t |t\rangle|\psi_0\rangle 
\sum_t' |t'\rangle|\psi^*_0\rangle$, easy to obtain from 
$|0\rangle|\psi_0\rangle |0\rangle|\psi^*_0\rangle$ by Hadamard gates.
Then one applies $U$ or $U^{*}$ in the same way as above
on each register $t$ or $t'$ times, 
by using conditional gates, in a manner similar to the one explained
in \cite{trace}.
We then do the QFT on both registers $|t\rangle$ and $|t'\rangle$,
giving $2^{-\nq}\sum_{\theta} a_{\theta} |\theta\rangle |\psi_{\theta}\rangle
\sum_{\theta'} a_{\theta'} |-\theta'\rangle|\psi^*_{\theta'}\rangle$,
where $\theta$ are the eigenphases of the unitary quantum map.
This corresponds to performing two phase estimations in parallel on
each copy of $|\psi\rangle$.  The functions $a_{\theta}$ are peaked
around the eigenvalues of $U$, with $|\psi_{\theta}\rangle$ the
corresponding eigenvector.
Then we use amplitude amplification to select the same eigenvalue
$\theta=-\theta'$.  This costs in general at most
$N\sqrt{N}$ operations in total.  It is possible
to improve this bound by taking advantage of the fact that
the wavefunction is  multifractal, as was the case for iterations.
For example in the simulation of (\ref{ISRM}), the wavefunction 
is multifractal in the $p$ representation.  So if one starts with
a basis vector in $p$ representation, the eigenvectors
will have a multifractal distribution on this particular basis vector.
Thus selecting $\theta=-\theta'$ costs a factor $N^{\alpha/2}$.
The selection of the diagonal part of the product of the two same
eigenvectors costs another $N^{\alpha/2}$.
This implies that the total
cost to get the same eigenvalue will be $N^{1+\alpha}$, as opposed
to $N^2$ for the classical algorithm.

Then the wavelet transform 
can be applied at a logarithmic cost.  This yields 
the numerator of the formula (\ref{Zarne}) in $N^{1+\alpha}$ operations.
The denominator can be obtained through classical approximations
with lower resolution, 
giving a total cost of $N^{1+\alpha}$ as opposed
to $N^2$ for the classical algorithm, but with the approximation that
the denominator corresponds to a lower resolution.
Then measurement of $a$ will give an histogram of 
$\sum_{b} |T_{|\Psi|^2}(a,b) |^2$
for the different values of $a$, from which the exponent can be
extracted.

The method described here enables to obtain the numerator in (\ref{Zarne}) more
efficiently on a quantum computer than on a classical device.
Nevertheless, this procedure needs to use a low-resolution
approximation of the denominator in (\ref{Zarne}) which will be
obtained on a classical device.  We have made numerical experiments
to see if computing classically the denominator with twice less
qubits (thus keeping the polynomial gain for the whole procedure)
leads to a good approximation of the $\tau_q$.  However, we were not
able to reach a large enough number of qubits to have sensible
data with half the qubits, and thus the evidence was inconclusive
(data not shown), although we believe the procedure should give a better
approximation to $\tau_q$ than with a computation of both numerator
and denominators at low resolution.  To circumvent this
problem, it is worthwile to investigate if a modification of the partition 
function (\ref{Zarne}) would be more adapted to quantum computation.

\subsection{Wavelet transform of $\psi$}

\begin{figure}[htbp]
\begin{center}
\includegraphics[width=1.0\linewidth]{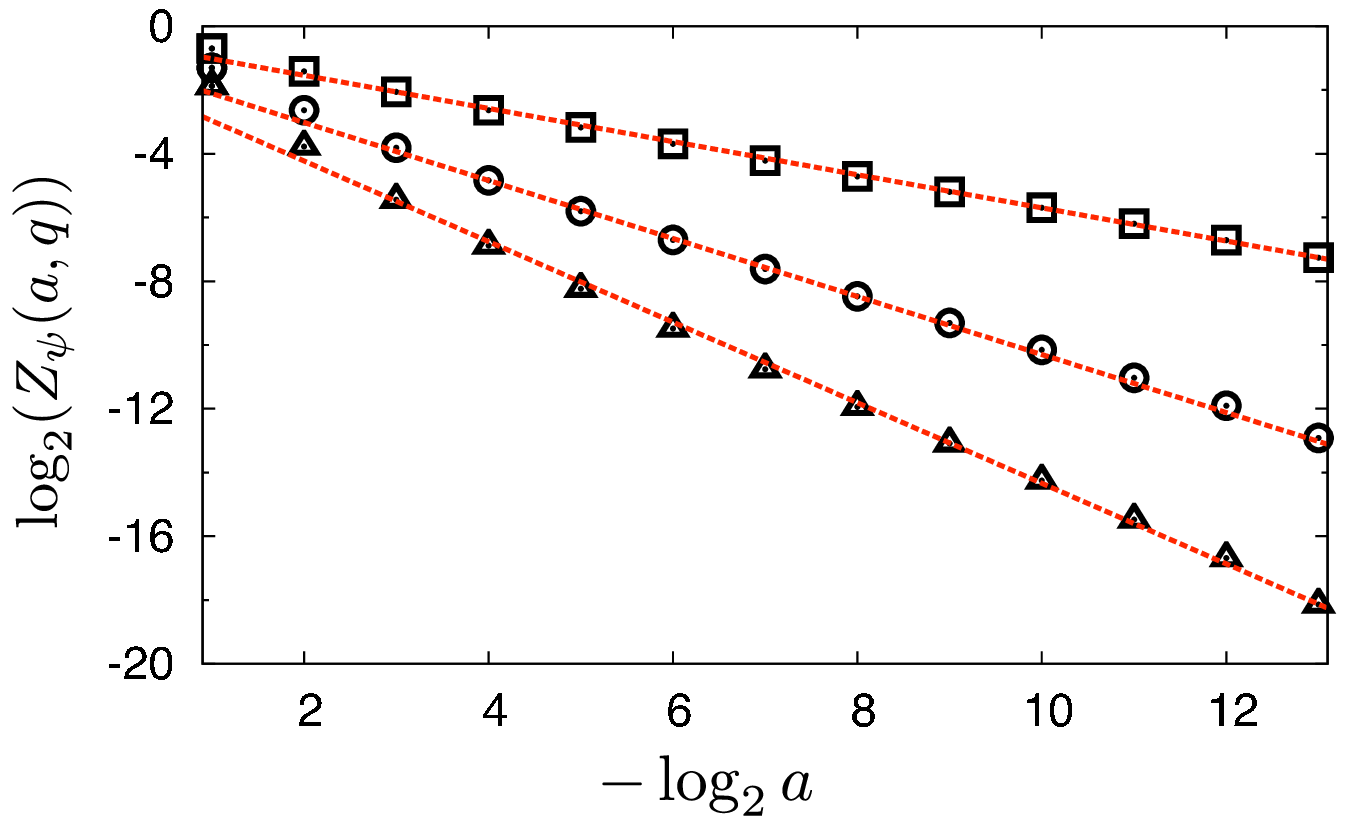}
\caption{(Color online)
Average of $\log_2 Z_{\psi}(a,q)$ 
as a function of the scale 
$a$ for $q=2$ ($\square$), $q=3$ ($\circ$), $q=4$ ($\triangle$). The  
slope of the fitting straight lines (dashed) gives  
$-\tau_q$. Averaging is done over $\sim 25000$
eigenvectors of \eqref{ISRM} of size $N=2^{14}$
for $n_2=3$ ($\gamma=1/3$).}
\label{fig3}
\end{center}
\end{figure}

In the preceding algorithm, one had to approximate the denominator in 
(\ref{Zarne}) through low resolution classical computation.  This entails an 
approximation of the multifractal exponent which can be difficult
to control.  In order to circumvent this difficulty, we studied the accuracy
of a different formula for the partition function 
which is more efficient to implement on a 
quantum computer. This is a generalization of the partition function 
\eqref{Zarne} which involves the wavelet transform of $\psi$ instead of
$|\psi|^2$, namely
\begin{equation}
\label{Zeff}
Z_{\psi}(a,q)= \sum_{b\in {\cal L}(a)}\left[\frac{|T_{\psi}(a,b)|^2}{\sum _{b\in {\cal L}(a)}|T_{\psi}(a,b)|^2}\right]^q.
\end{equation}
The advantage of such a choice for the partition function  is that
the QWT can now be performed directly on a state
$\sum_{i=0}^{N-1} \psi_i |i\rangle$, without using amplitude
amplification to build the $ |\psi_i|^2$ on the registers.
This new partition function $Z_{\psi}$ actually also has a scaling
$a^{\tau_q}$ for $a\rightarrow 0$ as in \eqref{eq:tauq}, as illustrated by
Fig.\ref{fig3}. 

\begin{figure}[htbp]
\begin{center}
\includegraphics[width=1.0\linewidth]{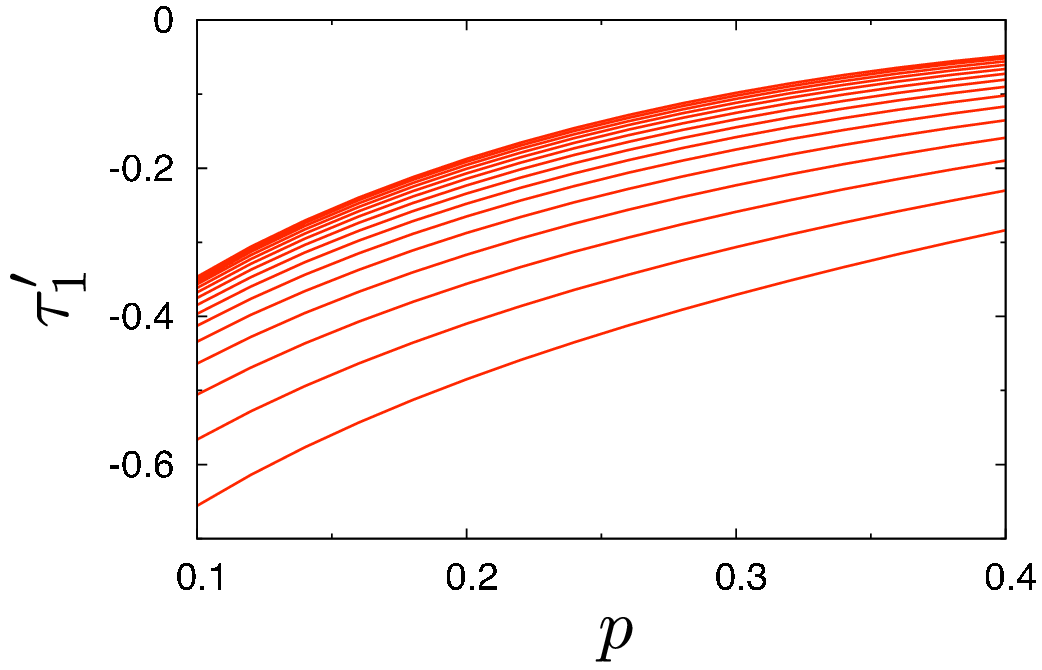}
\caption{(Color online)
$\tau'_1$ as a function of $p$ for the multifractal cascade;
vectors have size $N=2^n$ with
$n$ varying from top to bottom from $n=6$ to $n=20$.}
\label{fig4}
\end{center}
\end{figure}

\begin{figure}[htbp]
\begin{center}
\includegraphics[width=1.0\linewidth]{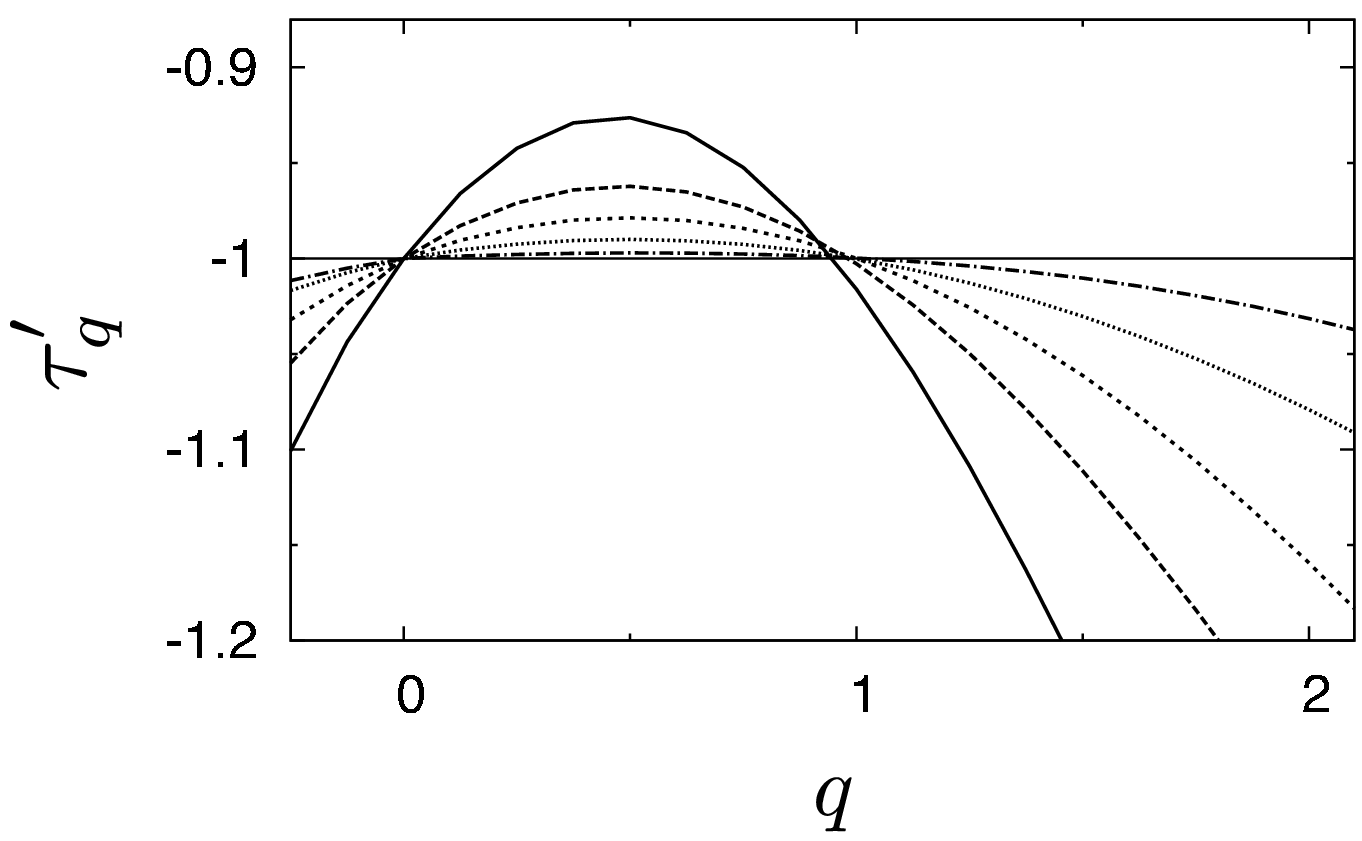}
\caption{$\tau'_q$ for eigenvectors of \eqref{ISRM}
as a function of $q$ for $n_2=3$ (solid), $n_2=5$  
(dashed), $n_2=7$ (dotted), $n_2=11$ (small-dot), and $n_2=13$ (dash-dot)
($n_1=1$). Average is done over $\sim 25000$ eigenvectors
of size $2^{12}$.}
\label{fig5}
\end{center}
\end{figure}

In fact, the quantities $\sum_{b\in {\cal L}(a)}|T_{\psi}(a,b)|^{2q}$ appearing
in the numerator of Eq.~\eqref{Zeff} should themselves contain some
information about fractality of the state $\psi$. Let us denote
$\tau_q'$ the corresponding multifractal exponents, that is the exponents
extracted from the partition function \eqref{Zeff} without the normalization,
that is from the scaling
\begin{equation}
\sum_{b\in {\cal L}(a)}|T_{\psi}(a,b)|^{2q} \sim a^{\tau'_q}.
\end{equation}
It might seem at first sight that the computation of $\tau'_1$ through QWT
could yield an exponential gain over classical computation, 
provided exponentially fast quantum
simulation of the state is possible. Indeed, for iterations of simple 
vectors (such as basis vectors) by an efficiently simulable
quantum map such as (\ref{ISRM}), the whole process
of iterations of the map and QWT is exponentially fast,
and the asymptotic behaviour of
$\sum_{b\in {\cal L}(a)}|T_{\psi}(a,b)|^2$ can be estimated also
exponentially fast by measuring $a$ and making an histogram of the
probabilities to extract the exponent $\tau'_1$. 
For the multifractal cascade, the slopes seem to converge to
a precise value which depends on the fractality
(see Fig.\ref{fig4}). 
However,
our numerical data show that for the quantum map (\ref{ISRM}) 
 the asymptotic behaviour of $\sum_{b\in {\cal L}(a)}|T_{\psi}(a,b)|^2$ is
independent of the system and seems universal. This can be seen in 
Fig.\ref{fig5}, where the multifractal exponents $\tau'_q$ extracted
from $\sum_{b\in {\cal L}(a)}|T_{\psi}(a,b)|^{2q}$ are shown.
It therefore seems that $\sum_{b\in {\cal L}(a)}|T_{\psi}(a,b)|^2$,
which can be obtained exponentially fast, does not
yield useful information in the case corresponding to actual
quantum simulation.  We shall return to this
quantity in Subsection C.  

In the remaining part 
of this subsection, we study the quantum simulation of the
quantity $Z_{\psi}(a,q)$, again for $q=2$. We show that it can be 
obtained with polynomial gain over classical computation. Since
the denominator of (\ref{Zeff}) is obtained exponentially fast, 
being just $\sum_{b\in {\cal L}(a)}|T_{\psi}(a,b)|^2$, the exponent
$\tau_2$ is now obtained without resorting to approximations.
\begin{figure}[htbp]
\begin{center}
\includegraphics[width=1.0\linewidth]{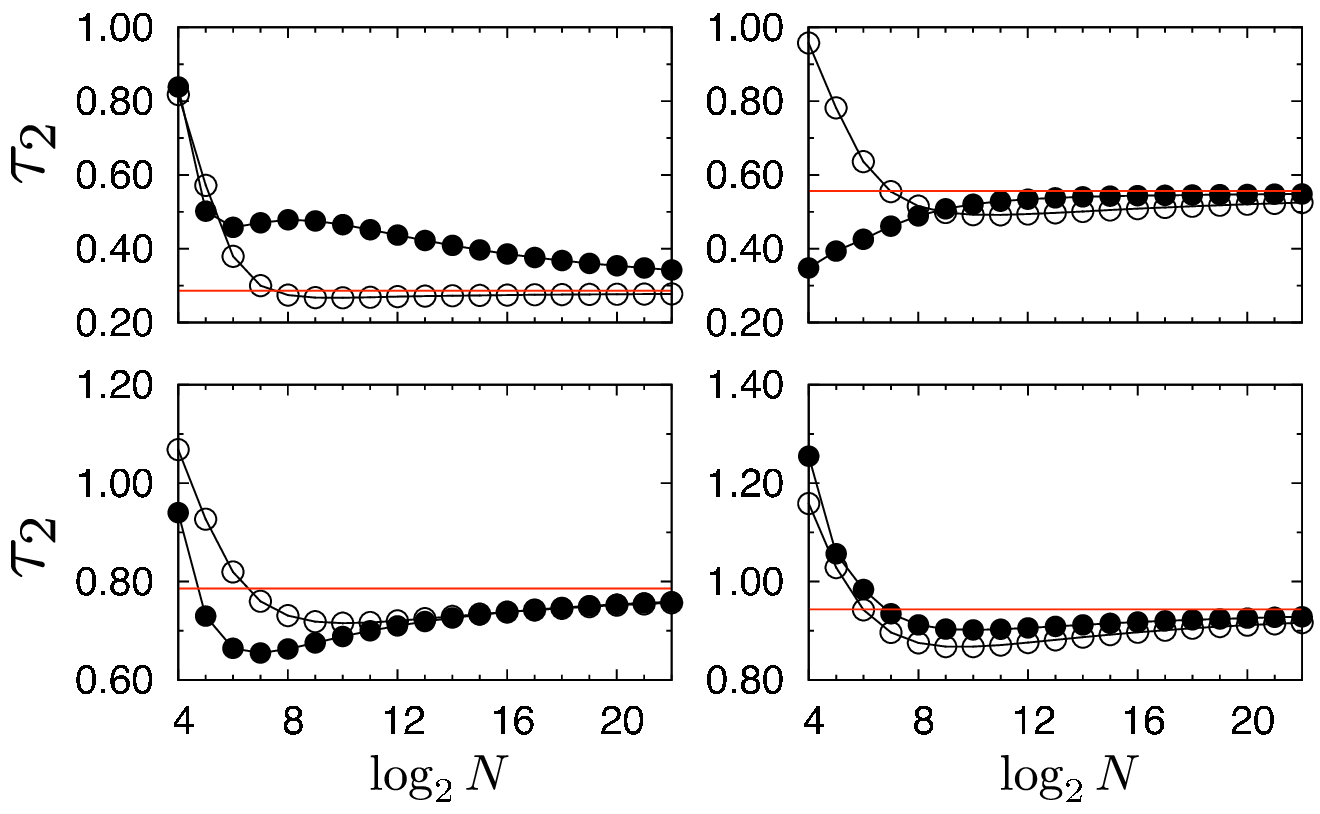}
\caption{(Color online) $\tau_2$ for the multifractal cascade computed with 
 $Z_{|\psi|^2}(a,q)$ (\ref{Zarne})
  (filled circles) and $Z_{\psi}(a,q)$ (\ref{Zeff})(empty circles).
 The values of $p$ are from left to right and top to bottom,
$p=0.1,0.2,0.3,0.4$. The straight horizontal lines are the theoretical values 
$\tau_2=-\log_2(p^2+(1-p)^2)$.
\label{fig6}}
\end{center}
\end{figure}

\begin{figure}[htbp]
\begin{center}
\includegraphics[width=1.0\linewidth]{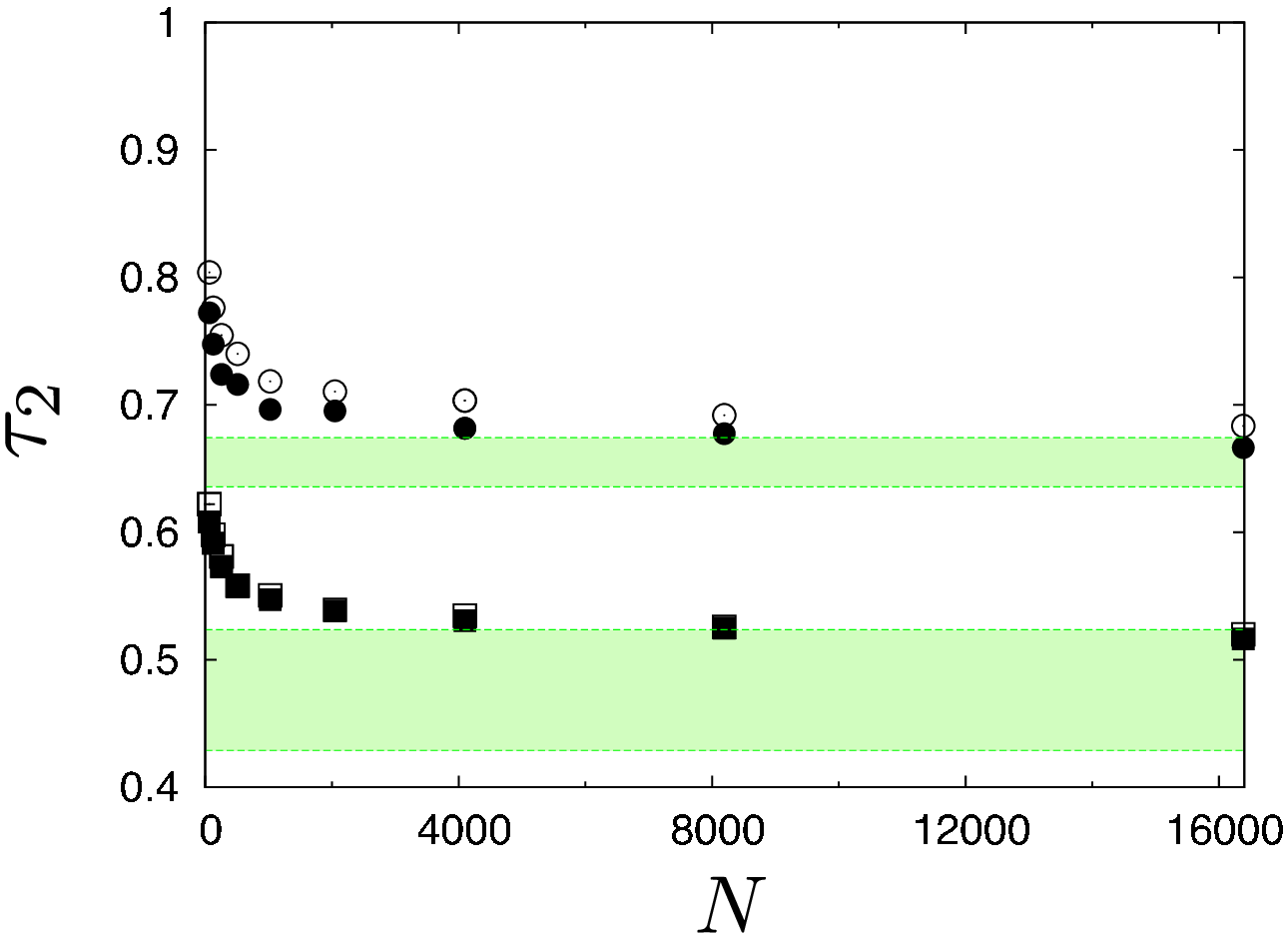}
\caption{(Color online) $\tau_2$ 
for eigenvectors of the intermediate map \eqref{ISRM} computed with 
   $Z_{|\psi|^2}(a,q)$ (\ref{Zarne})
  (filled symbols) 
  and $Z_{\psi}(a,q)$ (\ref{Zeff}) (empty symbols) for $n_2=3$ (squares) 
 and $n_2=5$ (circles) ($n_1=1$). Average is done over
$\sim 25\,000$ eigenvectors. The green/grey
  area shows the corresponding range of values obtained in
  \cite{martin} (top $n_2=5$, down $n_2=3$).}
\label{fig7}
\end{center}
\end{figure}
\begin{figure}[htbp]
\begin{center}
\includegraphics[width=1.0\linewidth]{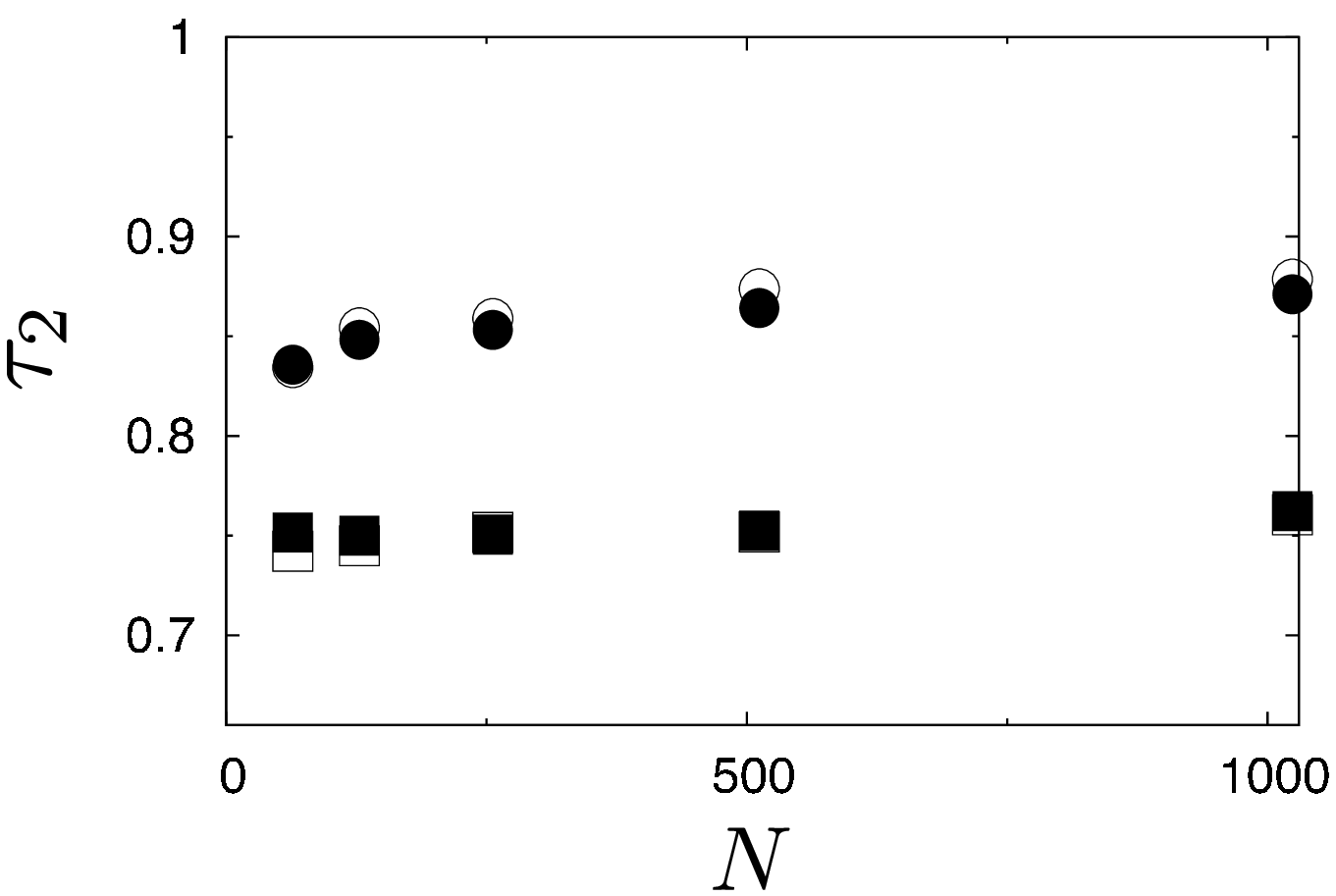}
\caption{$\tau_2$ for iterates of column vectors corresponding to
time $t=1000$ of the
 intermediate map \eqref{ISRM} computed with 
   $Z_{|\psi|^2}(a,q)$ (\ref{Zarne})
  (filled symbols) 
  and $Z_{\psi}(a,q)$ (\ref{Zeff}) (empty symbols) for $n_2=3$ (squares) 
 and $n_2=5$ (circles) ($n_1=1$). The
  number of vectors averaged for each case is $\sim 25\,000$.}
\label{fig8}
\end{center}
\end{figure}

Figs.\ref{fig6},\ref{fig7},\ref{fig8} show the behaviour of the numerical 
$\tau_2$
obtained from  (\ref{Zeff}) for increasing system sizes,
plotted together with the one obtained from (\ref{Zarne}). One
sees that both converge to the same value for large system size.
In the case of the multifractal cascade (Fig.\ref{fig6}), 
they both converge to 
the analytical value given by Eq.~\eqref{Dq_cascade}, 
while for eigenvectors of the intermediate
quantum map (\ref{ISRM}) (Fig.\ref{fig7}) the asymptotic
value is within the uncertainty
of the different numerical methods used in \cite{martin}. For iterates of 
column vectors of (\ref{ISRM}) (Fig.\ref{fig8}), one also obtains similar results 
with both methods.
This seems to indicate that this new partition function gives
asymptotically the same result as (\ref{Zarne}), and can be used
reliably to obtain this multifractal exponent.

To get the quantities $\sum_{b\in {\cal L}(a)}|T_{\psi}(a,b)|^4$, the
numerator of (\ref{Zeff}) for $Z_{\psi}(a,2)$, one needs to build the state
$\sum_{a,b} |T_{\psi}(a,b)|^2|a,b\rangle$.  This can be done through a 
procedure similar to the one exposed in Subsection A.  One starts
from two iterations of the map, followed by two QWT. This yields the state 
\begin{equation}
\sum_{a,b} T_{\psi}(a,b)|a,b\rangle\sum_{a',b'} T_{\psi}(a',b')|a',b'\rangle.
\nonumber
\end{equation}
Then amplitude amplification can be used to select the part $a=a'$ and
$b=b'$ in this expression, leading in at most $\sqrt{N}$ operations
to the state $\sum_{a,b} |T_{\psi}(a,b)|^2|a,b\rangle$.
From this state, measurement of register $\ket{a}$ yields a value of $a$ with
probability $\sum_{b\in {\cal L}(a)}|T_{\psi}(a,b)|^4$; an histogram
of these probabilities thus yields the exponent. 
In the case of the iterations of vectors through a quantum map
such as (\ref{ISRM}), the total cost of the method 
is at most of order $t\sqrt{N}$ operations, 
to be compared to $tN$ for iterating the vector classically up to time $t$.
Actually the scaling of the second moment of the wavelet transform vector 
$T_{\psi}$ has an exponent $\beta$ smaller than 1 
which means that $T_{\psi}$ is also fractal. This gives an actual quantum
cost of $tN^{\beta/2}$, where $\beta<1$ for fractal wavelet transform.

For eigenvectors of a quantum map, the same reasoning as above leads
to a quantum algorithm costing of the order $N^{1+\alpha/2+\beta/2}$ 
operations, as opposed
to $N^2$ for the classical algorithm, where $\alpha<1$ for fractal
wavefunctions
and $\beta < 1$ for fractal wavelet transform.

\subsection{Possibility of exponential gain}

\begin{figure}[htbp]
\begin{center}
\includegraphics[width=1.0\linewidth]{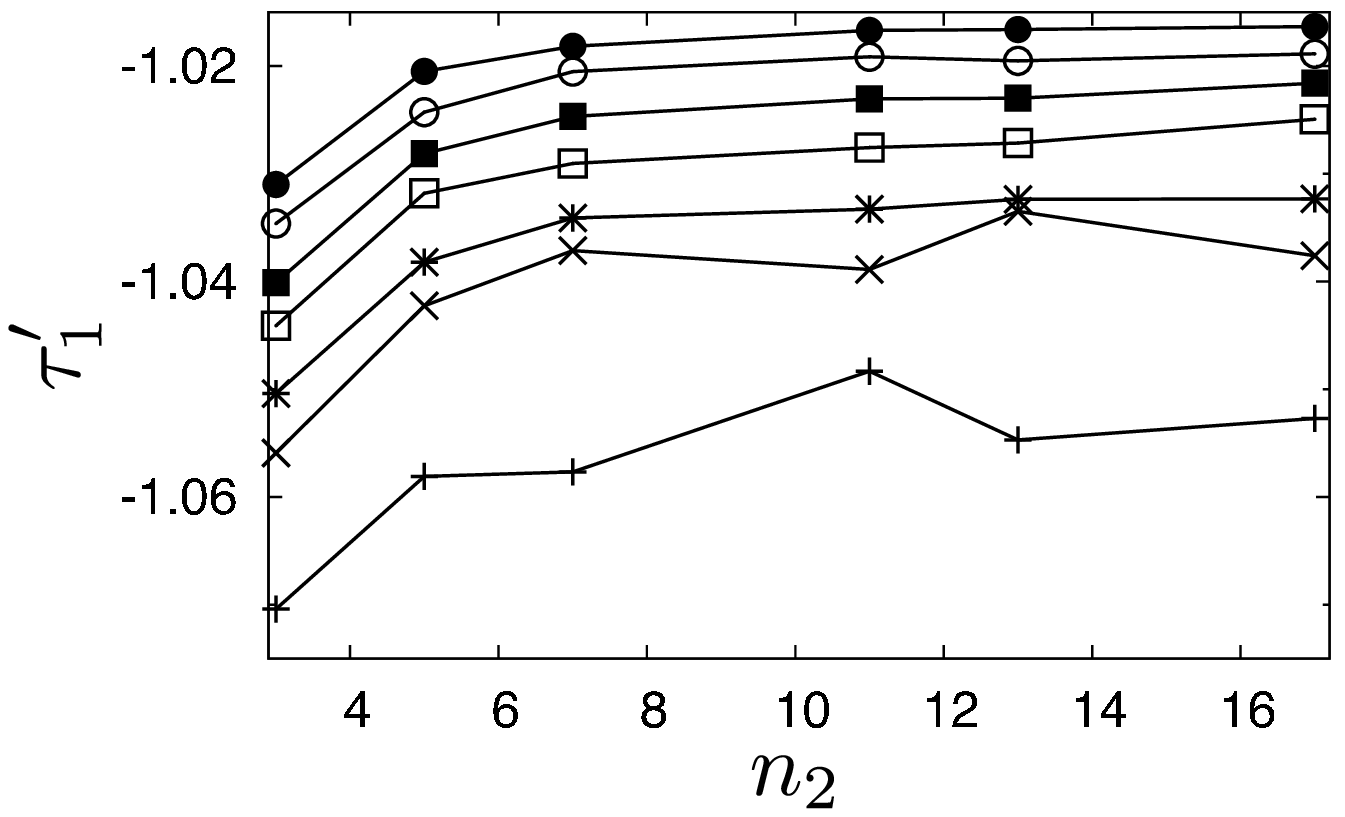}
\caption{$\tau'_1=\sum_b |T_\psi|^2$ 
for eigenvectors of the intermediate map \eqref{ISRM}
as a function of $n_2$ (with $n_1=1$)
for different values of $\nq$ (number of qubits) 
bottom to top $\nq=6,7,8,9,10,11,12$. The
  number of vectors averaged for each case is $\sim 25\,000$.}
\label{fig9}
\end{center}
\end{figure}

We have seen at the beginning of Subsection B that 
$\sum_{b\in {\cal L}(a)}|T_{\psi}(a,b)|^2$ can be estimated 
exponentially fast on a quantum computer.  The corresponding
exponent gave useful information for the cascade
(see Fig.\ref{fig4}), but
in the case of the quantum map
\eqref{ISRM}, our
simulations showed that the numerically extracted exponent 
converges for large system size to the same value, whatever the
fractality of the system (see Fig.\ref{fig5}).  Nevertheless, we present
numerical data on Fig.\ref{fig9} showing that for the multifractal
quantum map (\ref{ISRM}), although all exponents converge to the same
value, the rate of convergence seems to depend on the fractality
of the system.  If these numerical
indications correspond to a generic phenomenon for
multifractal quantum systems, then a rough
estimate of the degree of fractality of the system could
be obtained exponentially fast compared to classical algorithms.

\section{conclusion}

In this paper we have investigated the use of the quantum wavelet transform to 
extract information (namely the multifractal exponents) from a quantum
wavefunction produced by an efficient quantum simulation on a quantum
computer.  We have shown that various partition functions which can
be extracted from a wavefunction indeed enable to obtain the multifractal
exponents. Unfortunately, the partition function which can be obtained 
exponentially fast does not yield useful results, although our
numerical results indicate that it can give exponentially fast
a rough idea of the degree of fractality in a system.  We have
shown that other partition functions can be extracted, with a
polynomial gain compared to classical computation.

Our results indicate that the quantum wavelet transform can be applied 
efficiently to extract useful information for certain types 
of quantum simulation.  They show that once the full process of
simulation and measurement is taken into account, quantum simulations
of multifractal quantum systems yield polynomially efficient quantum
algorithms, which may be exponential in certain cases.  Our results
also show that the quantum wavelet transform is an effective tool
which acts in a complementary way from the more usual
quantum Fourier transform, enabling to extract certain types of
information efficiently.

We thank Marcello Terraneo and John Martin for helpful discussions, 
CalMiP
for access to their supercomputers, and
the French ANR
(project INFOSYSQQ) and the IST-FET program of the EC
(project EUROSQIP) for funding.


\end{document}